# A Flat-Panel Brightness Model for the Starlink Satellites and Measurement of their Absolute Visual Magnitude

Anthony Mallama

14012 Lancaster, Bowie, MD, 20715, USA

anthony.mallama@gmail.com

Version 2

2020 March 17

**Abstract**

The Starlink satellites are shaped like flat panels. The flat sides face zenith and nadir during normal operations. Their brightness is determined by the product of the solar illumination on the downward facing side of the panel multiplied by the area of that side projected toward the observer on Earth. This geometry leads to a unique brightness function that is not shared by other satellites. For example, the observed brightness is very sensitive to the solar elevation angle. There are circumstances where sunlight only illuminates the upward facing side of the satellite rendering it invisible to Earth-based observers. A brightness model depending on the solar aspect and the observer aspect of the flat panel, in addition to the satellite distance, is described. Absolute brightness is the only free parameter of the model, and it is taken to be that at a distance of 1,000 km when the solar and observer factors are unity. This model has been successfully fitted to a set of observed magnitudes. The absolute visual magnitude of a Starlink satellite as determined from this fitting is 4.1 +/- 0.1. The model could be used to determine the absolute magnitude of the Starlink satellite known as Dark Sat which has a special low-albedo coating.

**Introduction**

Thousands of internet communication satellites are currently being launched into low-earth orbit by the SpaceX company. The number and brightness of these Starlink satellites is a concern to astronomers because they will interfere with ground-based observations (Gallozzi et al., 2020, Hainaut and Williams, 2020, and McDowell, 2020). The amount of interference will depend partly upon the brightness of the satellites.

This paper outlines a model for computing the brightness of Starlink satellites. The model differs from general purpose formulas for celestial object brightness. General models rely on the phase angle which is the arc distance between the Sun and the observer as measured at the body.

The phase angle works well for the brightness of round bodies like planets (Mallama et al., 2017) because it is a convenient parameter for computing the illuminated-and-visible fraction of a sphere. Phase angle also works for some satellites but it does not apply to Starlink satellites. These bodies are flat and thin like a table top. The flat side with phased-array antennas, faces downward toward the Earth for communication. The opposite side, where the solar array is mounted, faces away from the Earth.

The brightness model described here is the product of the solar illumination on the downward facing side of the satellite multiplied by the projected area of that side as seen by the observer on Earth. That product is divided by the square of the distance from the observer to the satellite. Absolute calibration is derived empirically from observed magnitudes.

**Required Input Data**

The data values needed for a brightness prediction include positions of the satellite, the Sun and the observer. For the purpose of computing the illumination of the satellite panel by the Sun, the celestial coordinates for both of those bodies are needed. The satellite data values can be obtained from two-line ephemerides (TLEs). The solar positions can be taken from on-line ephemerides or can be computed. Additionally, the terrestrial coordinates of the observer will be needed to compute the projected area of the satellite that is seen.

**Observer Aspect**

The angular area of the satellite projected toward the observer is proportional to the cosine of the observer aspect angle shown as *c* in Figure 1. For example, when the satellite is at zenith *c* is 0.0 and cosine *c* is 1.0 because the flat side is projected entirely toward the observer.

Side *C* of the triangle is the Earth's radius, taken to be 6368 km. Side *A* is the distance from the geocenter to the satellite. For Starlink satellites at a nominal altitude of 550 km this distance is 6918 km. Angle *a* is 90 degrees plus the elevation of the satellite. Since, the angle *a* and its corresponding side *A* are known, angle *c* can be derived from the length of side *C* using the sine formula (see the figure). The area of the flat side of the satellite projected to the observer is proportional to the cosine of *c*. This is the projected profile, called *V*.

The angular area on the sky is also proportional to the square of the distance from the observer to the satellite, *B*. Angle *b* is known because *a*, *b* and *c* must sum to 180 degrees. Therefore, the length of side *B* can be determined at once from the sine formula. The angular area of the satellite projected to the observer, *M*, is proportional to the product of the projected profile, described above, divided by the square of the distance, or

$$M = V / B^2 = \cos(c) / B^2$$

Equation 1

The absolute value of the projected area (for example, square degrees) is not required as discussed in the sections which follow.

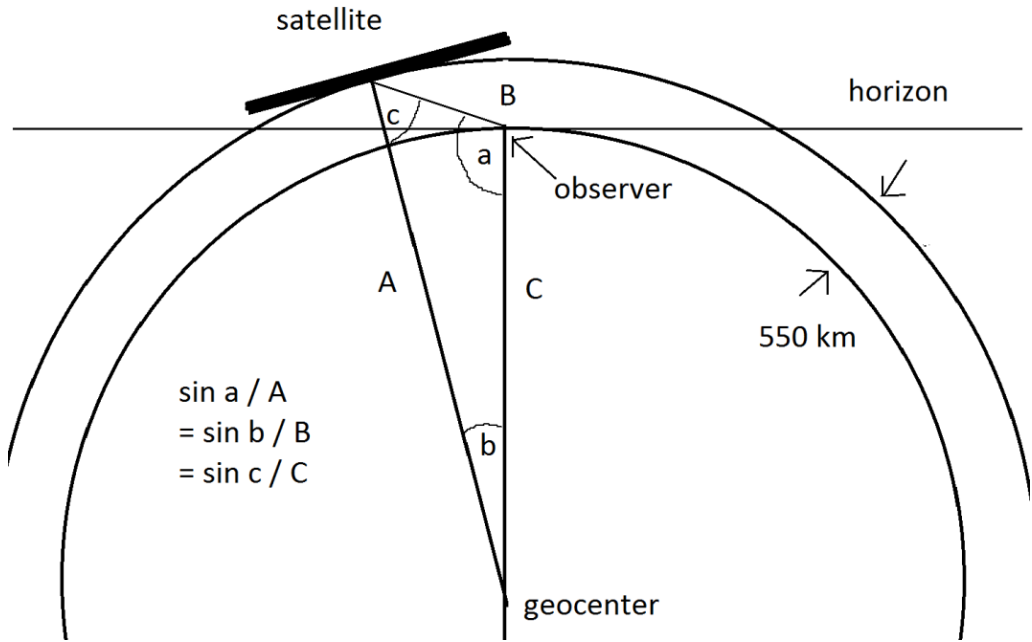

Figure 1. Observer aspect

**Solar Aspect**

The illumination, *S*, of the satellite's flat side is proportional to the cosine of the angle between the direction to the Sun and the perpendicular to the plane of that side, as shown in Figure 2. The perpendicular to the plane is in the same direction as is the satellite from the geocenter because the flat surface of the Starlink satellite faces down to nadir.

The direction to the Sun can be taken from on-line ephemerides (for example, JPL's Horizons) or computed from common astronomical formulas, as mentioned in the section on Required Input Data. When those two directions are known the angle between them can be computed using the cosine formula, and *S* is equal to that cosine. The cosine formula in terms of right ascension, *R*, and declination, *D*, of the Sun (R1, D1) and satellite (R2, D2) is

$$S = ( \sin (D1) * \sin (D2) ) + ( \cos (D1) * \cos (D2) * \cos ( R1 - R2 ) )$$

*Equation 2*

The Illumination of the nadir-pointing side of the satellite is the supplementary angle to that shown in the figure and its cosine is negative *S*. Furthermore, illumination is inversely proportional to the square of the distance to the Sun, *P*. So, the solar Illumination factor, *N*, is

$$N = -S/P^2$$

Equation 3

However, the solar distance only changes by a few percent over the course of the year. Since the effect of those changes on brightness are much smaller than the accuracy being sought from this model they are omitted here. Furthermore, the absolute value of the solar aspect (for example, watts) is not required. So,

$$N = -S$$

*Equation 4*

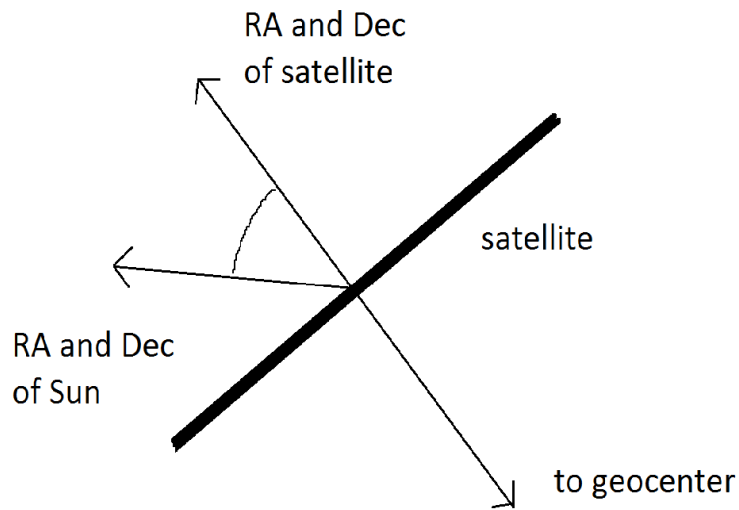

Figure 2. Solar aspect.

**Relative Brightness**

The previous two sections described the solar illumination aspect and the observer aspect of the brightness model. The product of these aspects divided by the square of the distance is proportional to the brightness of a Starlink satellite.

Some physical quantities, such as the flux of the Sun and the dimensions of the satellites were ignored. Therefore, predictions based on the solar and observer aspects alone are relative as opposed to absolute.

Figure 3 shows the relative brightness of Starlink satellites in the case where the azimuth of the satellite and the Sun are the same or where they differ by 180 degrees. There are three plots corresponding to solar elevation angles of -10, -20 and -30 degrees. Each of the plots illustrates brightness as a function of satellite elevation. The greatest brightness overall is for the Sun at -20 degrees elevation and the satellite near zenith. For small sun-ward elevations the satellites are brightest when the solar elevation is -30 degrees, however the satellites are eclipsed by the Earth's shadow when they exceed 31 degrees. When the solar elevation is -10 degrees the satellite is sunlit until its just 13 degrees from the horizon opposite the Sun.

The strong dependence of satellite brightness on solar elevation is a noteworthy feature of the flat-panel model and one which distinguishes it from a phase angle formula. For example, the brightness ratio shown in Figure 3 for a satellite at zenith with the Sun at -20 and -10 degrees elevation is 2.0, while for the phase angle representation this same ratio is only 1.1.

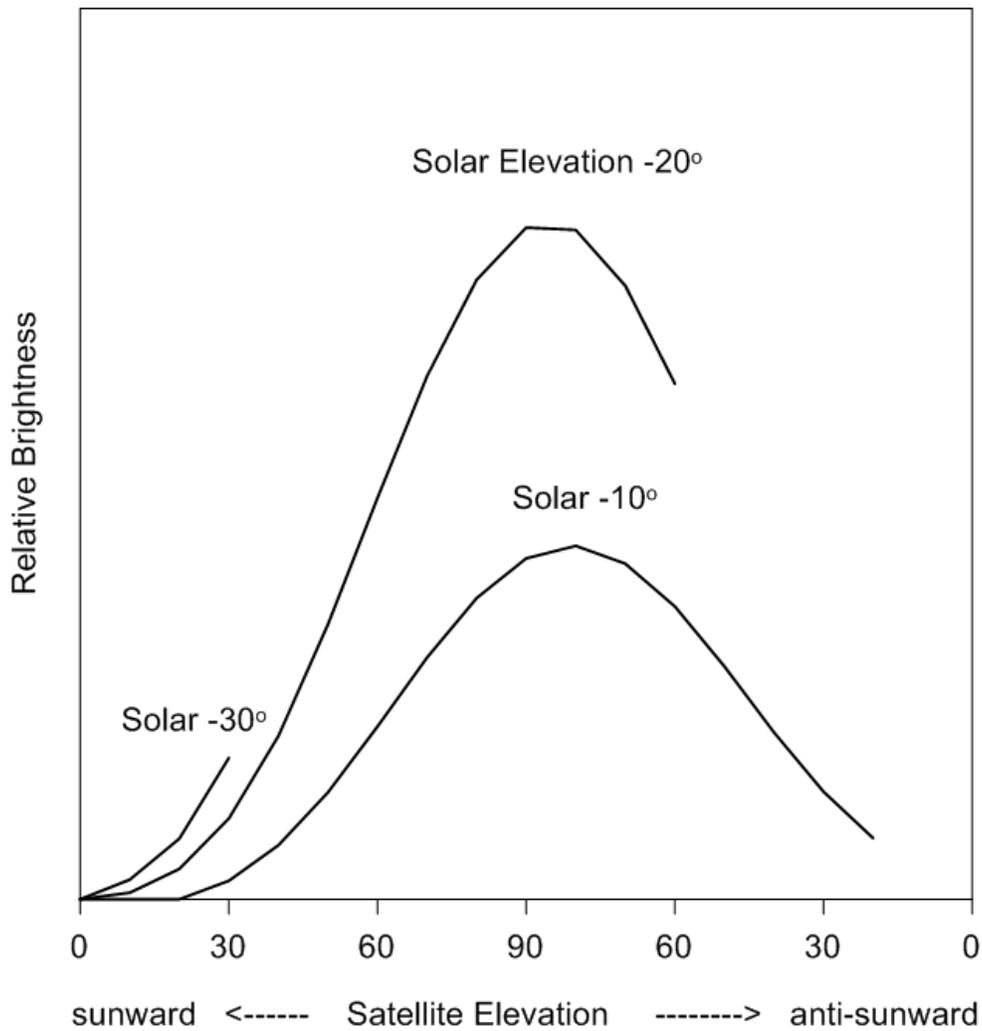

Figure 3. Relative brightness from the flat-panel model.

**Magnitude Calibration**

In order to predict the actual brightness of Starlink satellites seen in the sky, it is necessary to calibrate the relative brightness model using observations. The author obtained the magnitude estimates listed in Table 1 by visually comparing the brightness of 14 satellite passes to nearby stars with know magnitudes. The Earth coordinates of the observations are 38.982 N, 76.763 E and 43 m. "Cat" in the table refers to the Spacetrack catalog.

Table 1. Observed Magnitudes

```
Cat #     Cat Name        UTC Date      UTC Time  Magnitude
44747     STARLINK-1042   2020-Feb-27   23:45:00   5.2
44761     STARLINK-1056   2020-Feb-27   23:50:00   5.6
44759     STARLINK-1054   2020-Feb-27   23:55:00   5.8
44767     STARLINK-1062   2020-Feb-28   00:00:00   6.4
44758     STARLINK-1053   2020-Feb-28   00:05:00   5.9
44766     STARLINK-1061   2020-Feb-28   00:10:00   6.6
44771     STARLINK-1067   2020-Feb-28   00:15:00   6.5
44763     STARLINK-1058   2020-Feb-28   00:20:00   6.8
44761     STARLINK-1056   2020-Mar-15   23:59:30   4.5
44759     STARLINK-1054   2020-Mar-16   00:04:30   4.7
44767     STARLINK-1062   2020-Mar-16   00:10:00   5.0
44758     STARLINK-1053   2020-Mar-16   00:15:00   5.5
44766     STARLINK-1061   2020-Mar-16   00:20:00   5.5
44763     STARLINK-1058   2020-Mar-16   00:30:30   6.1
```

A modeled magnitude was derived from a modeled brightness value that corresponds to the satellite-Sun-observer geometry at the time of each observation. Absolute satellite brightness is commonly referenced to a distance of 1,000 km from the observer. So, $B$ in Equation 1 is in units of 1,000 km for the purpose of computing the modeled magnitudes. The average of the differences between the observed (O) and calculated (C) magnitudes is the calibration constant and it is equal to 4.1. This value is the brightness that a Starlink satellite would exhibit if it were 1,000 km from the observer, face-on to the observer and face-on to the Sun. Thus, the calibrated brightness model expressed in magnitudes is

$$M_V = 4.1 - 2.5 \log_{10} ( M N )$$

Equation 5

where $M_V$ indicates that the magnitude is in the visual band-pass.

Hainaut and Williams (2020) give the visible magnitude of a Starlink satellite at zenith and at a distance of 550 km as 4.2. However, they assume a spherical satellite modeled with a phase angle function evaluated at 90 degrees. Thus, their result cannot be readily compared to the absolute magnitude reported here.

**Error Estimation**

The characteristic difference between observed and modeled magnitudes is given by the standard deviation of the O-C values of the fit which is 0.3 magnitude. The accuracy of the absolute magnitude is the standard deviation of the mean which is 0.1 magnitude.

The visual magnitudes are estimated to be accurate to about 0.25 magnitude. Since the 0.3 magnitude standard deviation of the O-Cs includes this observational scatter, the fit to more accurate observations would be better. If the 0.25 magnitude observational accuracy is taken as a component of the 0.3 magnitude standard deviation in a statistical variance sense, then the accuracy of the model in the absence of observational error is reduced to about 0.2 magnitude.

**Satellites Illuminated But Not Visible**

A satellite may be in sunlight and above the horizon but still not be visible. Figure 4 shows the tilt of the satellite's flat dimension relative to the observer's horizon (for example, 10.1 degrees at satellite elevation 20 degrees). When that tilt exceeds the angular distance of the Sun below the horizon, the downward facing side of the satellite is unlit by the Sun if the two bodies are at similar azimuths from the observer. This renders the satellite invisible.

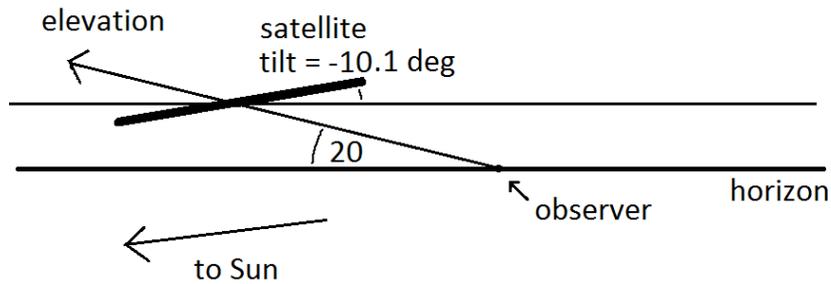

Figure 4. Backlit condition.

**Dark Sat**

SpaceX launched one satellite with a special low albedo coating designed to make it dimmer and thus less of a problem for ground-based astronomical observations. This satellite is numbered 44932 and named Starlink-1130 in the Spacetrack catalog, but it is commonly known as Dark Sat. A few visual observations of Dark Sat have been reported. However, the circumstances (solar aspect, observer aspect and distance) of these observations are unknown. Future observations that include the exact circumstances could be processed with the model described here in order to solve for the absolute magnitude of Dark Sat. Comparison with the average absolute magnitude of other Starlink satellites from visual observationindicate the effectiveness of the low albedo coating in that band-pass.

Tregloan-Reed et al. (2020) observed Dark Sat on the Sloan photometric system. They determined that the coating reduced the its reflectivity by 55% in the g' band-pass. They accounted for satellite distance in their study but neglected the solar and observer aspects.

**Limitations, Assumptions and Approximations**

The Starlink satellites have large solar arrays which rotate on their long axis to track the Sun. These arrays are not modeled here (which is a limitation) but there are three mitigating factors. One is that the arrays are attached to the top sides of the satellite bodies and will be partly hidden from ground-based observers. Another is that the panels are dark in order to absorb sunlight. The third is that when the phase angle of the satellite exceeds 90 degrees the panels are backlit and will not contribute any brightness. However, there will be times when the glass-coating on the panels reflect sunlight specularly to the observer which will produce a temporary brightness increase.

The model presented here assumes that the flat sides of the satellites reflect sunlight diffusely and isotropically. However, it is possible that the panels preferentially scatter sunlight in the forward or backward direction.

Another assumption is that the flat sides of the satellites are oriented directly downward in the nadir direction. They may sometimes be directed off-nadir especially prior to the satellites reaching their operational altitude.

The solar direction computed from the geocenter is not exactly the same as that from the satellite. This error is very small and it could be eliminated altogether by computing the solar direction from the satellite.

The Earth is assumed to be spherical. However, the globe is actually flattened by about one part in 300.

The observations in Table 1 are for satellites at the Starlink operational altitude of 550 km. Observations recorded for other satellites at lower altitudes shortly after launch were omitted because their orientation is not necessarily with the flat side facing nadir.

There are several selection effects in the observations. First is that the range of solar elevations is only from -10 to -17 degrees. Second is that satellite elevations are all at least 23 degrees. Third is that the observations were recorded when satellites were near the meridian. This simplified some aspects of the model computations which are being performed on a spreadsheet. Finally, one anomalously bright magnitude was omitted from the analysis. That brightness excess may have been due to a Sun glint.

**Discussion**

The model described here predicts the approximate brightness of Starlink satellites. The accuracy is estimated to be 0.3 magnitude or better. In practice, greater precision may not be very important to observational astronomers because any bright satellite can interfere with ground-based observations. So, the key concerns are to know the satellites' approximate magnitudes and their exact celestial coordinates.

Avoiding interference from Starlink satellites will require understanding the geometry of their eclipses. With that knowledge it is possible to identify areas of the sky where satellites will not interfere with observations because they are in the Earth's shadow. These eclipse areas will be small during twilight and large when the Sun is far below the horizon. This topic has been analysed by Gallozzi et al. (2020), Hainaut and Williams (2020) and McDowell (2020).

The model described here addresses the expected 1,584 Starlink satellites which are currently being placed into 550 km altitude orbits. Many thousands more will be put into orbits with altitudes ranging from 336 to 1325 km. The brightness model described here may be useful for those satellites depending upon their shapes and orientations.

**Conclusion**

A brightness model for the downward-facing side of Starlink satellites is described. Solar illumination on that side, the area of that side projected toward the observer on Earth and the satellite's distance contribute to its brightness. This geometry leads to a unique brightness function that is not shared by other satellites. For example, there are circumstances where sunlight only illuminates the upward facing side of the satellite rendering it invisible to Earth-based observers.

The only free parameter of the model is absolute brightness which is taken to be that at a distance of 1,000 km and when the solar aspect and the observer aspect are both unity. The model has been successfully fitted to a set of observed magnitudes. The absolute visual magnitude of a Starlink satellite as determined from fitting the model to the observations is 4.1 +/-0.1. The model could be used to determine the absolute magnitude of the Starlink satellite known as Dark Sat which has a special low-albedo coating.